# Energy-Aware Forwarding Strategy for Metro Ethernet Networks

Rihab Maaloul
National school of engineering,
LETI laboratory,
Sfax, Tunisia

mre7ab@gmail.com

Lamia Chaari
National school of engineering,
LETI laboratory,
Sfax, Tunisia

Lamia.Chaari@enis.rnu.tn

Bernard Cousin
University of Rennes 1
IRISA,
Rennes, France

bernard.cousin@irisa.fr

*Abstract*—**Energy optimization has become a crucial issue in the realm of ICT. This paper addresses the problem of energy consumption in a Metro Ethernet network. Ethernet technology deployments have been increasing tremendously because of their simplicity and low cost. However, much research remains to be conducted to address energy efficiency in Ethernet networks. In this paper, we propose a novel Energy Aware Forwarding Strategy for Metro Ethernet networks based on a modification of the Internet Energy Aware Routing (EAR) algorithm. Our contribution identifies the set of links to turn off and maintain links with minimum energy impact on the active state. Our proposed algorithm could be a superior choice for use in networks with low saturation, as it involves a tradeoff between maintaining good network performance and minimizing the active links in the network. Performance evaluation shows that, at medium load traffic, energy savings of 60% can be achieved. At high loads, energy savings of 40% can be achieved without affecting the network performance.**

## I. INTRODUCTION

Much research related to ICT has sought for more efficient solutions that can improve energy efficiency. The reduction of energy expenditure has become a major concern for telecommunications operators and Internet service providers. Currently, minimal research has addressed energy saving in Carrier Ethernet networks compared with those in IP networks. Metro Ethernet is the use of Carrier-Ethernet technology in Metropolitan Area Networks (MANs). It can be used to connect business local area networks (LANs) and individual end users to wide area networks (WAN) or to the Internet. Recently, significant innovations have been developed around Ethernet standards to meet the requirements of next generation broadband networks. These developments have made Ethernet a widely used technology, deployed at all levels of the network architecture (Access, Metro and Core networks).

Current Ethernet technologies rely on the Spanning Tree Protocol (STP), which was standardized in IEEE 802.1D [1], and its variants: Rapid Spanning Tree Protocol (RSTP) [2] and Multiple Spanning Tree Protocol (MSTP) [3]. These protocols manage the topology autonomously and provide a loop-free connectivity across a variety of network nodes. Although these protocols have been used for most Ethernet networks, they are not sufficiently powerful to satisfy Metro Ethernet network features as a Carrier-grade technology. These protocols have the following main shortcomings:

i) Inefficient use of resources: STP and its variants restrict the number of bridge ports being used, which reduces the available bandwidth, especially in cases of high load traffic.

ii) Suboptimal path: The path selection is based on a single spanning tree for the entire network (the shortest path tree roots at an arbitrary node) instead of the shortest path between source and destination node pairs.

iii) Re-convergence: STP implements a transactional distance-vector class of routing algorithm instead of a routing algorithm based on a network link topology database. This adversely impacts the convergence time of an Ethernet network after a topology change [4].

Recently, a new class of shortest path routing solutions has been introduced for Ethernet networks, the Shortest Path Bridging (SPB), standardized in IEEE 802.1aq [5].

SPB aims to ensure frame forwarding on the shortest path within a Shortest Path Tree (SPT) region of a network by using an extension of the IS-IS link state routing protocol [6]. In this way, SPB uses IS-IS procedures to construct and update the link state database in each SPT bridge.

Our work aims to develop an energy-saving strategy within a Metro Ethernet network. This idea is inspired from the EAR (Energy Aware Routing for Green OSPF) approach [7], which is designed for IP networks and is OSPF compliant. The EAR approach is an energy-saving strategy that is based on powering off parts of network devices (links and interfaces). Because we focus on Metro Ethernet networks, we propose a Metro Ethernet Energy Aware Forwarding Strategy (MEEAFS) that is IS-IS compliant.

OSPF [8] and IS-IS [9] are link state protocols that use Dijkstra's algorithm for computing the shortest path between node pairs. OSPF is an IP routing protocol only, while IS-IS supports the handling of MAC addresses; it is able to run directly over Ethernet as it is not tight to IP.

Since we focus on Metro Ethernet networks, we propose an energy-aware forwarding strategy for green carrier-Ethernet networks that is SPB-based and is IS-IS compliant.

The rest of this paper is organized as follows: Section II presents the main related works; Section III presents the problem formulation and subsequently describes both EAR and MEEAFS strategies. Section IV provides a performance study and evaluation for both EAR and MEEAFS strategies. Finally, we conclude this paper and we present our suggestions for future work in Section V.

## II. RELATED WORKS

In the literature addressing energy conservation, many strategies have been published. However, minimal research has focused on energy saving in Metro Ethernet networks. Besides, industrial efforts are devoted to Metro Ethernet, but none of them cares about energy economy at routing level. In order to better understand the possible ways to limit energy consumption in Metro Ethernet systems, we chose to overview the main approaches related to energy saving on IP networks that could be adapted to Metro Ethernet systems. In other words, we overview IP approaches which propose energy-aware routing protocols compatible with SPB-based Metro-Ethernet.

The authors in [10] have proposed an optimization model based on the traditional Multiple Spanning Tree Protocols (MSTP) green routing protocol. This model is intended for minimizing the energy consumption of Carrier-Ethernet networks. This optimization is performed in such a way that a portion of the network is forced by the objective function of the model to remain unused, thus making it possible to turn off the elements of that portion of the network. These network components are put into sleep mode to conserve energy. The main shortcoming of this approach is the use of MSTP, which is inefficient in the Metro Ethernet context.

In [11], the authors proposed an IP-related approach that switches the router to sleep mode during low-traffic periods and returns them to the working state during peak hours. This approach could be adapted to an Ethernet Bridge; however, this approach puts the whole router to sleep instead of powering off some of its interfaces (links), which leads to poor network performance.

In [7], the authors propose an Energy Aware Routing algorithm to power off a maximum of active links by dividing the network routers into three subsets (exporter, importer, neutral). The main idea of this algorithm is that only a subset of routers are elected to serve as exporters. Elected exporter nodes must have a high number of neighbors, so node election is based on the node degree. In that case, each exporter computes its SPT to export it toward its direct neighbors' routers. The latter, called importer routers, utilize the SPT of the associated exporter routers, but use the importer router as the root node. Doing this allows the powering off of the links that are no longer in the SPT of the importer routers. However, this algorithm considers neither the QoS constraints nor the traffic demand. Motivated by this EAR idea, we propose an energy-saving strategy applied within a Metro Ethernet network, considering a new criterion to select adequate exporter bridges and supporting acceptable network performance. To achieve this goal, we formulate an optimization model for the choice of exporter bridges that takes into account energy consumption impact. In our model, an energy consumption function needs to be minimized that is subject to a set of constraints involving the minimal performance guarantees, which are explained in the next section.

## III. PROBLEM FORMULATION AND ENERGY-AWARE STRATEGIES

To achieve energy conservation, the Network Management System (NMS) is designed to manage and solve an optimization problem that considers the network topology and traffic demand [12]. Our work takes into account the NMS considerations as inputs of our model. It additionally aims to power off the maximum number of links while retaining sufficient bandwidth on residual paths. In this section, we present the problem formulation and the EAR strategy.

### A. Problem Formulation

TABLE I. SUMMARY OF NOTATION

| Variable | Description |
|---|---|
| G (N, E, W) | Directed graph where N is the set of nodes, E is the set of edges between two nodes, and W is the set of weights associated with each arc. An arc is equivalent to a directed edge. |
| \|N\|,\|E\| | Cardinality of set N and E, respectively. |
| $\varepsilon(i,j)$ | The energy consumption of the edge (i, j) ∈E. |
| $c(i,j)$ | The capacity of the edge (i, j) ∈E. |
| A | The set of arcs that are directed links between nodes. |
| D | A set of all traffic demands. |
| $d^{st}$ | Demand of traffic flow from $s$ to $t$. |
| $f_{ij}^{st}$ | Traffic demand from $s$ to $t$ that traverses the arc from $i$ to $j$. |
| $f_{ij}$ | Traffic routed through the link from $i$ to $j$. |
| Es | The set of links utilized to route traffic. |
| Th | Threshold of link load. |
| $\mu_{i,j}$ | The maximum link utilization. |

Consider a Carrier-Ethernet core network presented as a weighted graph G (N, E). The nodes in N represent bridges, and the edges in E represent connections between those bridges. Let |N| and |E| be the number of network nodes and links, respectively. Each link (i, j) ∈ E between two nodes i, j ∈ N has an energy consumption ε(i, j) and a capacity c(i, j).

The traffic demand between a pair of nodes could be presented as $d^{st}$, where s ∈ N is the originated node and t ∈ N is the destination node. $f_{ij}$ denotes the number of traffic units routed through the link from i to j.

The optimization of power consumption can be expressed formally with the following objective and constraints:

$$\text{Minimize} = \sum_{(i,j) \in A} x_{(i,j)} \, \varepsilon(i,j) \quad (1)$$

$$\text{with } x_{(i,j)} = \begin{matrix} 1 & if\ edge(i,j) \in E\ is\ switched\ on. \\ 0 & otherwise. \end{matrix}$$

Subject to:

$$\sum_{j=1}^{|N|} f_{ij}^{st} - \sum_{j=1}^{|N|} f_{ji}^{st} = \begin{matrix} d^{st}, & \forall i = s \\ -d^{st}, & \forall i = t \\ 0, & \forall i \neq s,t \end{matrix} \quad (2)$$

$$f_{ij} \leq \mu_{i,j} \times c(i,j) \text{ With } \mu_{i,j} \in ]0,1] \quad (3)$$

Equation (2) represents the classical flow conservation constraints ensuring that flows entering and leaving a node are equal. Equation (3) forces the link load to be smaller than the maximum target utilization $\mu_{i,j}$.

Let $SPT_k$ be the subgraph of G obtained by the $k^{th}$ bridge using Dijkstra's shortest path first algorithm toward all network nodes. Let SPG (N, $E_s$) be the subgraph of G obtained by the superposition of all $SPT_k (k = 1,...,|N|)$, i.e. SPG = $\cup_{i=1..|N|} SPT_i$.

$E_s$ includes all of the links that belong to at least one $SPT_k$, and identifies all of the paths used to route traffic. We consider this type of link as an active link.

$$E_s = \text{link}(\cup_{i=1..|N|} SPT_i) \quad (4)$$

It can be demonstrated that

$$|E_s| \subseteq |E| \quad (5)$$

Proof: The equality between $|E|_s$ and $|E|$ holds when all link costs are equal. In this case, the routing paths correspond to the shortest paths, and SPG coincides with G because when all links have equal cost, the shortest path between two neighbor switches is always the direct link between these two neighbors.

In more general cases, when the link weights are different, the number of active links is smaller than $|E|$, i.e. $|E_s| \subset |E|$. A first step to obtain a reduction of energy consumption is to switch off the links belonging to the set $E - E_s$. The minimum value of $|E_s|$, i.e. the minimum number of unidirectional links needed to route traffic between any pair of bridges, is

$$L_{min} = 2(N - 1) \quad (6)$$

$L_{min}$ is the minimum number of links that guarantees total connectivity for the network. The condition (6) is verified when all of the nodes compute the same SPT. According to this condition, by switching off the $|E| - 2(|N| - 1)$ links of G, we would obtain the maximum energy savings, but this leads to traffic congestion and subsequently poor network performance.

According to the NMS, given the traffic demand and the network topology as inputs, the outputs of the optimization problem will be the set of links to switch off and the paths that the traffic should use over the residual links. Table I lists a summary of the parameter definitions.

### B. EAR description

The EAR algorithm [7] is a distributed, energy-aware routing protocol that is able to save energy by performing the election of exporter nodes. The scheme involves forcing a subset of routers to use some routes that are different from those elected in their SPTs. The set of network routers is divided into three subsets: exporters, importers, and neutral routers. This scheme is achieved by the three following phases: i) During the first phase (election of exporter routers), each node calculates its shortest path tree (via the Dijkstra method); ii) In the second phase, called Modified Path Tree (MPT) evaluation, every importer router fulfills its new path tree by using the associated exporter's tree and extracts the links to be switched off; iii) During the third phase, called routing path optimization, after removing the links that have been switched off, each router computes its paths, using the Dijkstra algorithm, on the residual network topology. The aim of this step is to update the routing paths and to ensure that all of the routers are on the same reference topology.

Our work aims to apply the EAR strategy to Metro Ethernet. Hence, our MEEAFS proposal computes an SPB to enhance Metro Ethernet performance. We also propose a new criterion for exporter bridge selection and energy conservation. Unlike the EAR algorithm, which is based on the node degree for the exporter router selection, MEEAFS is based on the link energy consumption of the nodes.

### C. MEEAFS algorithm

Our MEEAFS algorithm uses as input the graph model formulation that is explained above. However, given the tradeoff between energy saving and network capacity, MEEAFS offers two enhancement criteria. The first one is based on the link energy consumption $å(i,j)$ and the second one is based on the link capacities $c(i,j)$.

As in the MEEAFS algorithm, the set of network bridges is divided into three subsets: exporter (EB), importer (IB), and neutral bridges (NB). The MEEAFS algorithm can be summarized as follows:

*1) Selection of the EBs:* During this phase, each bridge computes its energy impact. This information is obtained by computing the energy impact of the line cards (EILCs) of each bridge. In [13], an energy profile is defined as the energy consumption (in Watt-hours) in the function of the traffic load and throughput of a particular network component. EILC can be estimated due to the knowledge of the topology and the traffic conditions by means of the SPB and IS-IS protocols. $W_i$ denotes the power consumption weight of bridge i and is defined as follows:

$$W_i = \sum_{k=1}^{nc_i} EILC_{i,k} \quad (7)$$

where $i \in N$ and $nc_i$ is the maximum number of cards in bridge i.

According to (7), the bridge power consumption expresses the energy impact as a function of link traffic load.

Bridges having the lowest power consumption are inserted in the exporter bridge list, called EB_list. The direct neighbors cannot be considered as candidate EBs. This process is applied recursively on the remaining bridges. Consequently, the bridges inserted into EB_list have the minimum energy consumption impact. The links associated to EBs are less likely to be switched off when an IB uses the SPT's EB as its own. In Figure 1(a), initially we assume that the A and B bridges are candidates to be exporters. Fig. 1(a) shows an example of a network graph with EILC weights. According to (5), $W_A = 0.3 + 0.4 + 0.3 + 0.4 = 1.4$ and $W_B = 0.3 + 0.3 + 0.4 + 0.5 = 1.5$. The bridge A is elected as EB, and hence B is an IB that will use A as its packet forwarder.

*2) Modification of the SPT of IB:* In this phase, each *IB* has to execute a slight translation of Dijkstra's algorithm based on its associated EB, in order to identify the set of links that can be switched off. Each IB transforms its SPT into a Modified Shortest Path Tree (MSPT).

As explained in the first section, SPB uses the IS-IS standard to construct and update a link state database in each bridge. Thus, the complexity of the classical SPB remains the same, when the IB computes an SPT in which the root node is the associated EB. Fig. 1(d) shows an example of a network graph in which A is elected as an EB and B is an IB that uses the SPT of A as its own modified SPT. We denote MSPT (B, A) the imported SPT of A for B. The bridge B has to force itself as the tree root node by changing the direction of the link between A and B.

Once all of the IBs have computed their MSPT, any network link that no longer appears in any MSPT will be declared as a link to turn off. Thus, each IB checks iteratively if a given link (that belongs to the links to turn off, denoted by $L_{off}$) can legitimately be turned off or must be kept in its modified shortest path tree (MSPT). For each iteration, the considered link is removed from the forwarding table if the link load $f_{ij}$ is smaller than a fixed threshold Th.

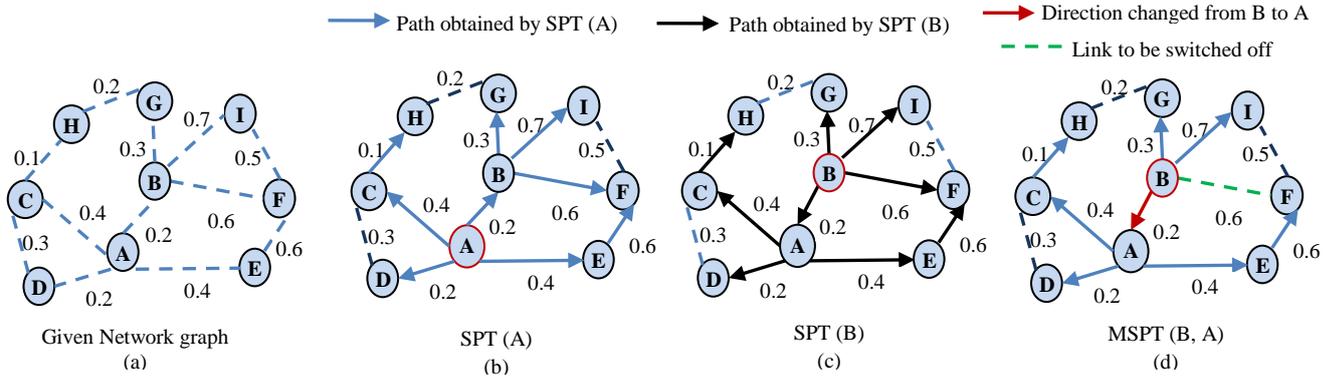

Fig. 1. Illustrative example for MEEAFS phases (1) and (2)

*3) Forwarding path optimization:* At the end of the previous phase, each IB has to assess the modified forwarding path tree MSPT. Thereafter, each IB indicates the set of links that has to be removed. In order to optimize the forwarding path trees, each IB having at least one switched off link processes IS-IS Hello until the topology database has been updated. Once the update process is terminated, SPB performs the shortest path calculation on the residual network topology.

**Algorithm 1.** *MEEAFS*

1. **Input**: a network graph $G(N, E)$
2. **for** i = 1; i<= N; i++ do
3. calculate the weight of each bridge using (7);
4. **end for**
5. B_list=list[N];
6. EB_list= ø;
7. IB_list=ø;
8. NB_list=ø;
9. **while** (B_list ! empty)
10. EB_list= EB_list ∪ finding_exporter(B_list);
11. **end while**
12. **while**(EB_list ! empty)
13. IB_list= IB_list ∪ finding_importer(B_list, EB_list);
14. **end while**
15. **while**(B_list ! empty)
16. NB_list= NB_list ∪ finding_neutral(B_list, EB_list, IB_list);
17. **end while**
18. $L_{off}$= ø; /*links to be turned off */
19. i=0;
20. **while**(IB_list ! empty)
21. modify_the_shortest_path_tree_of(IB_list[i]);
22. $L_{off}$=$L_{off}$ ∪ fixing_link(IB_list[i])
23. **if** (L - $L_{off}$= 2(B- 1)) according to (6)
24. exit and go to 28. ;
25. **end if**
26. i++;
27. **end while**
28. **for each** l∈$L_{off}$
29.   **if** $f_{ij}$> Th /* according to (3)*/
30. $L_{off}$ = $L_{off}$− l;
31. **end if**
32. **end for**
33. /* path optimization */
34. $E_s = E − L_{off}$;
35. compute _all_shortest_path using SPB for the residual topology $G'(N, E_s)$;
36. **output** $G'(N, E_s)$/* network graph with the set of links to be used*/.

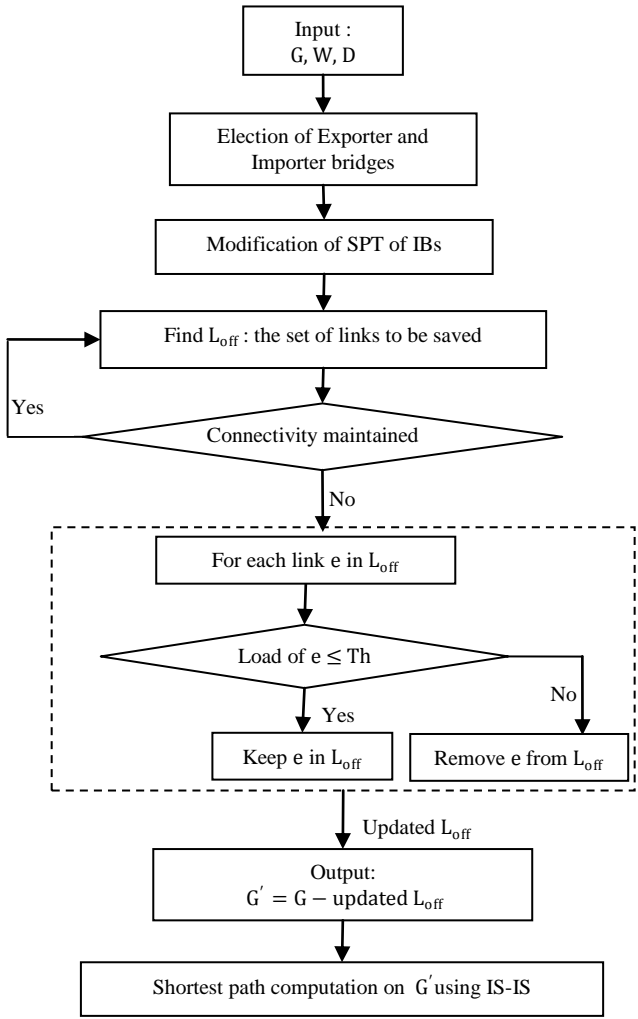

Fig. 2. Flowchart describing the operation of *MEEAFS*

## IV. PERFORMANCE ANALYSIS

This section presents the MEEAFS performance obtained in different scenarios using the ns-3 simulator. Initially, the simulator reads a weighted adjacency matrix of an input topology of 50, 100, 200 and 300 nodes, respectively. These weights represent the energy impact of links (two adjacent nodes). We have considered four core network topologies: the first one is composed of 50 nodes and 348 links; the second one composed of 100 nodes and 964 links; the third one is composed of 200 nodes and 1,926 links; and the fourth one is composed of 300 nodes and 2,276 links. Each bridge is assumed to generate traffic toward any other bridge. Traffic demands arrive at the network nodes following a Poisson process with arrival demands rate $\lambda$ and required $k_d$ traffic units that is randomly generated between $0.001*c$ and $0.1*c$ ($c$ being the link capacity in traffic unit). We consider the following two evaluations:

(1) A comparison of the MEEAFS algorithm versus the EAR algorithm with different topologies.
(2) A general performance analysis of the MEEAFS algorithm.

The obtained results are the average of ten independent runs. To evaluate the energy savings that could be achieved by EAR and MEEAFS, we consider the $\sigma$ index:

$$\sigma\% = 100 \cdot \frac{|E|-|E_s|}{|E|-L_{min}} \quad (8)$$

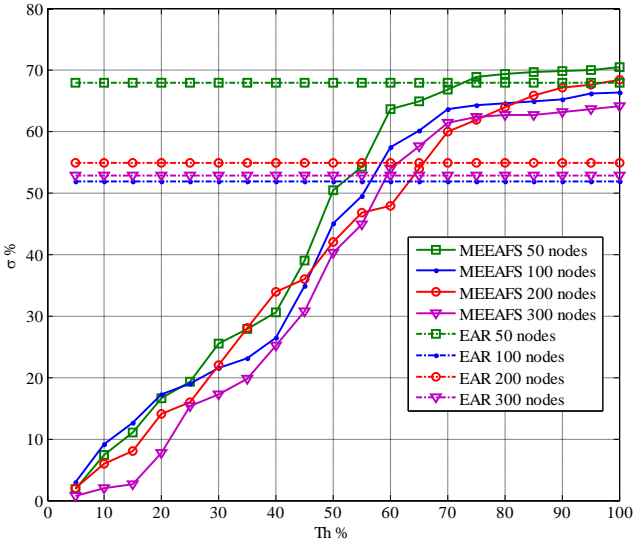

Fig. 3. Energy saving versus link load threshold

Figure 3 depicts the energy savings obtained by EAR and MEEAFS algorithms for the four topologies. The EAR energy savings are constant, because the EAR algorithm has not defined a threshold either to turn off or to keep links in the active state. The EAR algorithm also does not consider the traffic load.

Therefore, when EAR is used, 55% of possible links are turned off. However, we notice that the performance of MEEAFS is dependent on the Th value. The power saving obtained by MEEAFS increases with the increase of the threshold Th. This algorithm was able to achieve more than 65% energy savings when the value of the threshold was greater than 75%.

Figure 4 depicts the energy savings obtained by MEEAFS in both medium and high loads of traffic. We have already observed that in medium loads more energy saving can be achieved compared with high loads. Indeed, as the traffic load increases, fewer links can be turned off.

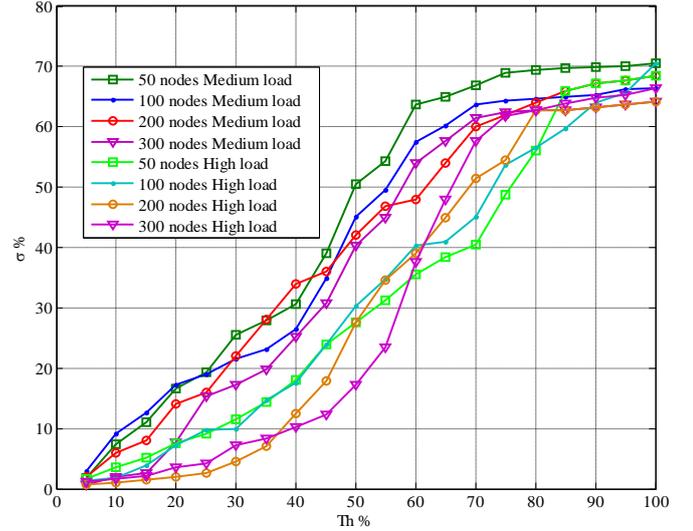

Fig. 4. Energy savings versus link load threshold

In order to analyze the impact of the MEEAFS algorithm on network performance, we evaluate the average traffic load on active links by varying the number of turned off links (thus varying Th). We introduce the $\rho\%$ parameter (average link load of active links), which is computed as follows [14]:

$$\rho\% = 100 \cdot \frac{\sum_{i=1}^{|E_s|} \rho_i}{|E_s|} \quad (9)$$

Where $\rho_i$ is the traffic utilization of link i.

In Fig.5, the average traffic load on active links as a function of Th is reported when MEEAFS is performed. In this scenario, we generate medium loads with traffic demand rate $\lambda$ equal to 0.2. We observe that our strategy achieves satisfactory results in terms of average link load.

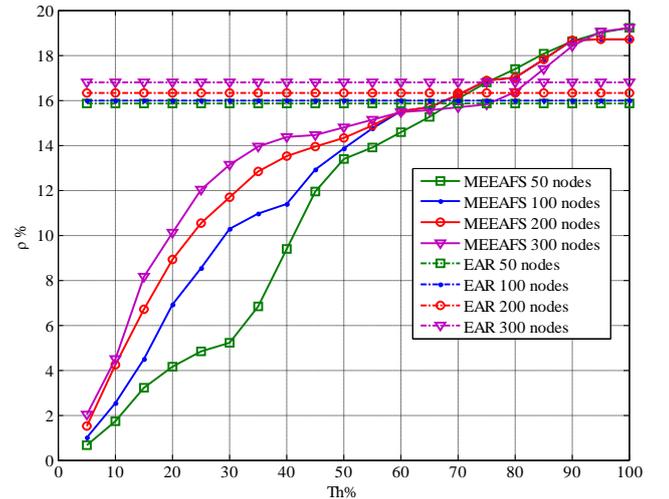

Fig. 5. Average traffic load in active links at medium load

Figure 6 reports the average traffic load on active links. In this scenario, we generate high loads with traffic demand rate $\lambda$ equal to 0.7. In this case, we observe that our strategy can achieve an acceptable link load if the threshold does not exceed 55%.

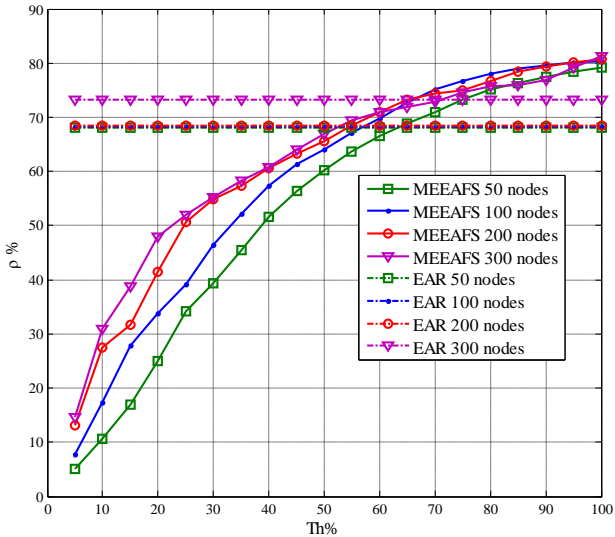

Fig. 6. Average traffic load in active links at high load

The increase of threshold value corresponds to the increase in energy savings. As shown in Fig. 3, the MEEAFS algorithm reports significant performance at medium traffic loads, with energy saving that can exceed 60%. Moreover, the MEEAFS algorithm can achieve approximately 40% energy savings at high traffic loads without affecting the network performance. Therefore, the choice of the threshold value in each scenario is critical to modulate the energy savings. Hence, MEEAFS could potentially guarantee a reduced impact on traffic performance.

Load balancing is considered to be a requirement that should be fulfilled in Carrier Ethernet. Hence, the third performance analysis is devoted to measuring the fairness of the traffic distribution on the active links $E_s$. The fairness index FI is used to measure whether the traffic load is fairly distributed among all of the links. We utilize Jain's Fairness Index [15]:

$$FI\% = \frac{\left(\sum_{i=1}^{|E_s|} \tilde{n}_i\right)^2}{|E_s|*\sum_{i=1}^{|E_s|} \tilde{n}_i^2} \quad (10)$$

When FI=1, this indicates that the traffic is distributed in a fair way.

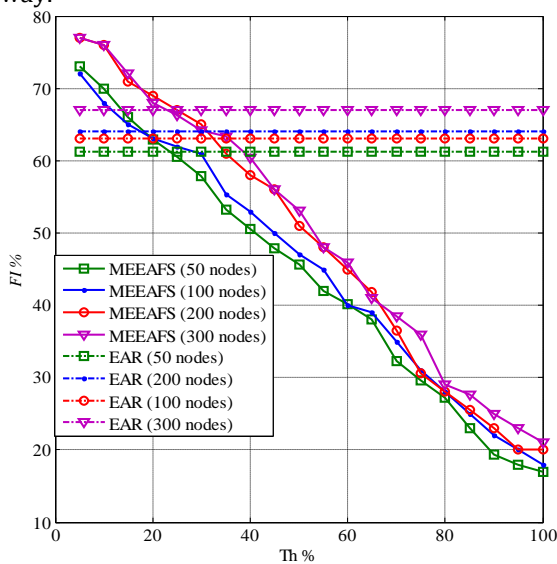

Fig. 7. Fairness index versus link load threshold at medium load

Figures 7 and 8 summarize the value of FI obtained by MEEAFS and EAR in medium and high loads of traffic, respectively.

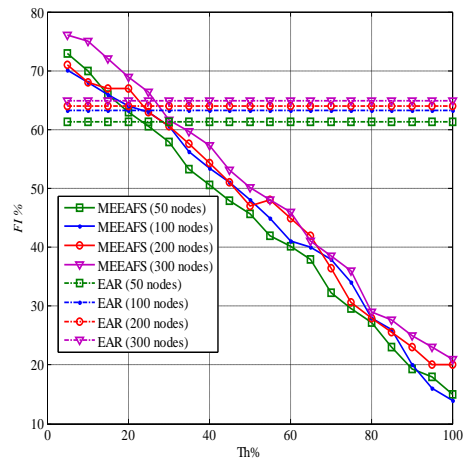

Fig. 8. Fairness index versus link load threshold at high load

We observe that the fairness index in both load conditions (high and medium) is nearly similar. Similarly, the fairness index decreases when the number of turned-off links increases, i.e., when the threshold Th values increase. Thus, as we discussed earlier, the choice of threshold value is critical to obtain the desired performance versus the energy saving gain.

V. CONCLUSION AND FUTURE WORK

This work proposes a new routing algorithm, called MEEAFS, to save energy in Carrier Ethernet networks. It allows a subset of bridge interfaces to be turned off. Our algorithm is based on a modification of the Internet Energy Aware Routing Algorithm [7]. Among the EAR algorithm's limitations, the traffic loads on links are ignored and the energy consumption impacts of link cards are not taken into account. Our algorithm resolves those constraints. It is based on a heuristic that identifies the exporter bridges and fixes a value of link load threshold to ensure acceptable network performance. Nevertheless, MEEAFS could affect the average route length. It is not evaluated in this paper and it will be considered in the next work.

The presented results show that important energy savings may be achieved with MEEAFS for our scenarios. This was typically true at medium traffic loads and for threshold values higher than 50%. However and obviously, at high traffic loads, MEEAFS cannot achieve significant energy reduction without degrading network performance.

We tested the behavior of MEEAFS by considering only the core network segment. Based on these encouraging results, our future work will consider an actual network topology, including all of the network segments: core, metro, and access. For example, realistic network topologies from available datasets might be used, such as the RocketFuel [16] and Topology Zoo datasets [17], and dynamic (cyclostationary) traffic matrices could be generated [18].